\documentstyle[sprocl]{article}
\def\be{\begin{equation}}
\def\ee{\end{equation}}
\def\bea{\begin{eqnarray}}
\def\eea{\end{eqnarray}}
\begin{document}

\title{A NOTE ON THE STABILITY OF QUANTUM SUPERMEMBRANES}

\author{MICHIO KAKU}

\address{Physics Dept., City College of the City University of New York\\
New York, N.Y. 10031, USA}


\maketitle\abstracts{
We re-examine the question of the stability of quantum supermembranes.
In the past, the instability of supermembranes was established by
using a regulator, i.e. approximating the membrane by $SU(N)$
super Yang-Mills theory
and letting $N \rightarrow \infty$. In this paper, we 
(a) show that the instability persists even if we directly
examine the continuum theory, which then opens the door to other
types of regularizations.
(b) give heuristic arguments that even a theory of unstable membranes at the 
Planck length may still be compatible with experiment.
(c) resolve a certain puzzling discrepancy between earlier works on the stability
of supermembranes.}

\section{Quantum Supermembranes}

To analyze the quantum stability of supermembranes, we 
start with the Hamiltonian in the light cone gauge for the
supermembrane$^{1,2}$:

\begin {equation}
H = \int d ^ 2 \sigma \left [
{ 1\over 2 } ( P ^ I ) ^ 2 +
{ 1\over 4 }
( \{ X ^ I, X ^ J \} ) ^ 2 
- {i \over 2 } \bar \theta \Gamma ^ I \{ X ^ I  , \theta \} \right ]
\end {equation}
where $I = 1,2,...,9$ and $\{ A, B \} = \partial _ 1 A \, 
\partial _ 2 B - ( 1 \leftrightarrow 2)$.
To demonstrate the instability of this Hamiltonian without going to a
$SU(N)$ super Yang-Mills theory$^3$, we let $ f ( \sigma _ 1 , \sigma _2 )$
represent a function of the membrane world-variables and consider
$X _ \mu ( f)$, which describes a string-like configuration. 
The potential function in the Hamiltonian vanishes if we substitute
this string-like configuration. In other words, along these string-like
configurations, the wave function can \lq\lq leak" out to infinity, and hence the
system is unstable.

We first choose
variables. Let $a = 1,2, ... 8$, and $I = 1,2, ... 9$. Then choose
co-ordinates $ X ^ I = \left ( x _ 9, Y _ 9 , x _ a , Y _ a \right )$.
where we have split off the string co-ordinate by setting
$x = x ( f )$, while $Y$ cannot be written as a string.
We can fix the gauge by choosing $Y_9=0$.

Then the Hamiltonian can be split up into several pieces$^4$:

$H = H _ 1 + H _ 2 + H _ 3 +  H _ 4$
where:

\begin {eqnarray}
H _ 1  &= &- { 1 \over 2 }
\int d ^ 2 \sigma \, \left [
\left ( { \partial \over \partial x }\right) ^ 2
+
\left ( {\partial \over \partial x _ a } \right ) ^ 2 \right ]
\nonumber \\
H _ 2 &= &- { 1 \over 2 } \int d ^ 2\sigma \,
\left ( { \partial \over \partial Y _ a }\right) ^ 2 
+
{ 1 \over 2 }
\int d ^ 2 \sigma d ^ 2 \bar \sigma 
d ^ 2 \sigma ' \,
\left [
Y _ a ( \bar \sigma ) z ^ T ( \bar \sigma, \sigma ')
z ( \sigma ' , \sigma ) Y _ a ( \sigma ) \right ]
\nonumber \\
H _ 3 &= &- { i \over 2}
\int d ^ 2 \sigma  d ^ 2 \bar \sigma \,
\bar \theta ( \bar \sigma ) \left [
z ( \bar \sigma, \sigma ) \gamma _ 9 + 
z _ a ( \bar \sigma ,\sigma )\gamma _ a  \right] \theta ( \sigma )
\nonumber \\
z ( \bar \sigma , \sigma ) &=
& \delta ^ 2 ( \bar \sigma , \sigma ) \partial _ { \sigma _ 1 }x
\partial _ { \bar\sigma _ 2  } - ( 1 \leftrightarrow 2 )
\nonumber \\
z _ a ( \bar \sigma , \sigma ) &=
&
\delta ^ 2 ( \bar \sigma , \sigma ) \partial _ { \sigma _ 1 } x _ a 
\partial _ { \bar\sigma _ 2  } - ( 1 \leftrightarrow 2 )
\end {eqnarray}
and the index $\sigma$ is shorthand for $\{ \sigma _ 1 , \sigma _ 2 \}$.
Notice that $z(\bar \sigma, \sigma)$ is an anti-symmetric
function. 
$H _ 4 $ contains other terms in $Y$, which will vanish at the end of
the calculation.

Now consider the term $H _2 $. The eigenfunction
for $H_2$ is:

\begin {eqnarray}
\Phi _ 0 & = & A ( {\rm det } \, \Omega ) ^ 2 \exp
\left ( - { 1 \over 2 }
\int d ^ 2 \bar \sigma d ^ 2 \sigma \, Y _ a ( \bar \sigma )
\Omega ( \bar \sigma ,\sigma ) Y _ a ( \sigma ) \right)
\nonumber \\
H _ 2 \Phi  _ 0 & = & 4 \int d ^ 2 \sigma \, 
\int d ^ 2 \sigma '
\, z ^ T ( \bar \sigma , \sigma ' ) z ( \sigma ' , \sigma )
\, \Phi _ 0 
\end {eqnarray}

Let us
introduce eigenvectors $E _  { MN}^ \sigma$,
where $M \neq N$, as follows:

\begin {equation}
\int d ^ 2 \sigma \, z ( \bar \sigma , \sigma ) 
\,E _ { MN } ^ \sigma = i \lambda _ { MN } E_ { MN } ^ \sigma
\end {equation}
where $M,N$ label a complete set of orthonormal functions,
which can be either continuous or discrete, and $\lambda _ { MN}$
are the anti-symmetric
eigenvalues of $z$, so the eigenvalue of $H _ 2$ becomes:
$\sum _ { M,N } | \lambda _ { MN } |$.

Now change fermionic variables to:

\begin {equation}
\theta (\sigma ) = \sum _ { M\neq N} \theta ^ { MN } 
E _ { MN } ^ \sigma
\end {equation}

The original fermionic variables are real, so
$\theta ^ { MN\dagger} = \theta ^ { NM}$.

Then $H_3$ reduces to:

\begin {equation}
H _ 3  =  
\sum _ { M<N} \theta ^ { MN \dagger} 
\left ( \lambda _ { MN } \gamma  _ 9 +
\lambda _ { MN } ^ a \gamma _ a \right )
\theta ^ { MN} 
\end {equation}

After a bit of work, we find that:

\begin {equation}
H _ 3 \xi = - 8 \sum _ { M<N} \omega _ { MN }\xi
\end {equation}
where: $\omega _ { MN } = \{ ( \lambda _ { MN } ^ 2
+ ( \lambda _ { MN} ^ a ) ^ 2 \} ^ { 1/2}$.
To find total energy, we now sum the contributions of
$H _ 2$ and $H _3$:

\begin {equation}
( H _ 2 + H _ 3 ) \Psi = 8
\sum _ { M< N} \left ( 
|\lambda _ { MN }| - \omega _ { MN } \right ) \Psi
\end {equation}

Then we see that $\omega _ { MN}$
asymptotically approaches $|\lambda _ { MN }|$ in this limit,
so that the ground state energy of $H _ 2$ and $H _ 3$
vanishes.
In conclusion$^4$, we see that the principal contributions to the
ground state energy comes from $ H _ 2 $ and $H _ 3 $, which in turn
cancel if we are far from the membrane.
Furthermore, we see that the energy eigenvalue of the operator
is continuous for the ground state, which means that the system is
unstable.

Although the system is unstable, we speculate how this may still
be compatible with known phenomenology. 
We note that because 
the decay time of such a quantum membrane is on the order of
the Planck time, it is possible that unstable membranes decay
too rapidly to be detected by our instruments.
Consider the standard decay of the quark bound state:

\begin {equation}
\Gamma = { 16 \pi \alpha ^ 2 \over 3 }
{ \left | \psi ( 0 ) \right | ^ 2 \over
M ^ 2 }
\end {equation}
where $\phi (0 )$ is the wave function of the bound state at the origin,
and $M$ is the mass of the decay product. 

We expect that $| \psi ( 0 )|$ to be on the order of a ${\rm fermi}^ {-3}$,
the rough size of the quark-anti-quark bound state. For our purposes,
we assume that the membrane is on the order of the Planck length.
On dimensional grounds, we therefore expect that 
the lifetime of the membrane to be on the order of
$T \sim { M ^ 2 L ^ { -3} }$
where $L$ is the Planck length.

For relatively light-weight membranes, we find that the 
lifetime is much smaller than Planck times, so we will, as expected,
never see these particles. 

For very massive membranes,
we find that the lifetime becomes arbitrarily long, which seems
to violate experiment. 
However, the coupling of very massive membranes,
much heavier than the Planck mass,
is very small, and hence they barely couple 
to the particles we see in nature.
Again, we see that unstable membranes cannot be measured in the laboratory.

\section{Resolving a Discrepancy in the Stability of Membranes}
In this section, we try to resolve a certain discrepancy with regard to the
instability of supermembranes. In ref. 5, it was pointed out that it is 
possible to choose a gauge where the Hamiltonian becomes quadratic.
Thus, it appears that the non-linearity of supermembranes, and hence
their instability, is an illusion.

To resolve this puzzle, go to light cone co-ordinates: 
$\{ X ^ + , X ^ - , X _ { d-2} , X _ i \} $, where $i = 1,2,3, ... d-3$.
Our goal is to re-write everything in terms of $X _ i$.

Choose the gauge $ P ^ + = p ^ + ; \,\, X ^ + = p ^ + \tau$.
We still have one more gauge degree of freedom left. 
We choose ($\Delta$ is the volume term):
\begin{equation}
 P _ { d-2} ^ 2 + \Delta = \partial _ a X _ i \partial _ a X _ i
\end {equation}
This determines $ P _ { d-2}$ such that the left-hand side
is quartic, but the right hand side is quadratic.

Now let us solve the constraints. The $ P ^ 2 + \Delta$ constraint
can be solved for $ P ^ -$, which now becomes the new light cone Hamiltonian.
We find:
\begin {equation}
P ^ - = { 1 \over 2 p ^ + } \left ( P _ i ^ 2 + 
w ^ { ab} \partial _ a X _ i \partial _ b X _ i \right )
\end {equation}
Notice that the light cone Hamiltonian has now become quadratic!
And lastly, $X _ { d-2}$ is fixed by the other constraints $
\partial _ a X _ \mu P ^ \mu = 0$.

At this point, we now have a Hamiltonian $P ^ - $ defined entirely in terms
of quadratic functions, defined in terms of the canonical variables
$X _ i$, $P _ i$. The action seems to be stable.
But this contradicts all our previous results.

There is, however, a loophole to this proof. Notice that 
the $ \dot X _ \mu P ^ \mu$ term in the Lagrangian $L (  X, P )$
decomposes as
$\dot X  _ \mu P ^ \mu = 
{ p ^ + } P ^ - -  \dot X _ { d-2 } P _ { d-2}
- \dot X _ i P _ i$.

The key point in all of this is that $\dot X _ { d-2} P _ { d-2}$
does not vanish.
If we take the $\tau$ derivative of $X _ { d-2} $, we find that
this term contains non-linear functions of
$\dot X _ i$, and hence 
changes the canonical commutation relations between
the transverse variables. Thus, the non-linearity of the theory
creeps back in and is now hidden in the commutation relations.
This resolves the apparent discrepancy.

\end{document}